\documentclass[a4paper, 11pt]{article}
\pdfoutput=1

\usepackage{jcappub}
\usepackage{graphicx}
\usepackage{amsmath}

\title{
Dark Matter vs. Neutrinos: The effect of astrophysical uncertainties and timing information on the neutrino floor
}

\author{Jonathan H.~Davis}
\affiliation{Institut d'Astrophysique de Paris, 98 bis boulevard Arago, 75014 Paris, France}

 \emailAdd{jonathan.h.m.davis@gmail.com}

\abstract{
Future multi-tonne Direct Detection experiments will be sensitive to solar neutrino induced nuclear recoils which form an irreducible background to light Dark Matter searches. 
Indeed for masses around 6 GeV the spectra of neutrinos and Dark Matter are so similar that experiments are said to run into a neutrino floor, for which sensitivity increases only marginally with exposure past a certain cross section. In this work we show that this floor can be overcome using the different annual modulation expected from solar neutrinos and Dark Matter. Specifically for cross sections below the neutrino floor the DM signal is observable through a phase shift and a smaller amplitude for the time-dependent event rate. This allows the exclusion power to be improved by up to an order of magnitude for large exposures.
In addition we demonstrate that, using only spectral information, the neutrino floor exists over a wider mass range than has been previously shown, since the large uncertainties in the Dark Matter velocity distribution make the signal spectrum harder to distinguish from the neutrino background. However for most velocity distributions it can still be surpassed using timing information, and so the neutrino floor is not an absolute limit on the sensitivity of Direct Detection experiments.
}

\begin{document}
\maketitle
\flushbottom
\section{Introduction}
There is strong evidence that most of the mass in the Universe is composed of Dark Matter (DM), which interacts so weakly with normal matter that its existence can only be inferred by its gravitational effects~\cite{Ade:2013zuv}. Hence the composition of DM is unknown: specifically if DM is made of new fundamental particles it is not clear to what extent it interacts with the particles of the Standard Model.
One such location where DM exists in abundance is in our own galaxy (and indeed most others) where it is distributed approximately spherically around the galactic disc~\cite{Bertone:2004pz}. Direct Detection experiments aim to probe the mass of the DM and how it couples to nuclear matter, by searching for particles  of DM originating from this halo interacting with nuclei on Earth.

The state-of-the-art experiments which set the strongest limits on the spin independent DM-nucleon cross section are CRESST-II~\cite{Angloher:2014myn}, LUX~\cite{Akerib:2013tjd}, SuperCDMS~\cite{Agnese:2014aze} and XENON100~\cite{Aprile:2012_225}\footnote{Less constraining limits also exist from e.g. CDEX~\cite{Yue:2014qdu}, CDMS-II~\cite{Agnese:2013cvt}, CoGeNT~\cite{Davis:2014bla,Aalseth:2014jpa} and PandaX~\cite{Xiao:2014xyn}.}. With no current discovery signal of DM these experiments have set exclusion limits on the DM-nucleon cross section at the level of $10^{-40}$~cm$^2$ for DM with masses around 5-10~GeV (hereafter referred to as `light' DM) and at the level of $10^{-45}$~cm$^2$ for heavy DM around 50~GeV. The sensitivity of these experiments is limited by the rate of background nuclear recoils, which are predominantly due to mis-identified electronic-recoil events from beta or gamma radiation and neutrons from radioactive shielding materials. With improvements in shielding technology the rate of these backgrounds in state-of-the-art experiments is steadily decreasing.
Indeed provided the number of background events remains constant then the sensitivity of these experiments to smaller DM-nucleon cross section improves linearly with larger exposure (the product of detector mass and running time).

However with Direct Detection experiments in the future set to be up to several orders of magnitude larger (e.g. Darwin~\cite{2012JPhCS.375a2028B}, DEAP-3600~\cite{2012JPhCS.375a2027B}, LZ~\cite{2011arXiv1110.0103M}, SuperCDMS-SNO~\cite{cdms_sno_pres} and XENONnT~\cite{2012arXiv1206.6288A})  this trend will not continue unabated, due to an irreducible background from neutrino-induced nuclear recoils~\cite{Billard:2013qya,Monroe:2007xp}. Since neutrinos can not be shielded against it is impossible to keep the number of background events small for large exposures.
Indeed the era of large neutrino backgrounds for light DM searches will soon be upon us: as has been shown previously (see e.g.~\cite{Grothaus:2014hja,Billard:2013qya,Ruppin:2014bra,Monroe:2007xp}) a low-threshold xenon-based experiment with a mass of $\sim 4$~kg will detect on average one nuclear-recoil event due to solar neutrinos per year (primarily those from the $^8$B chain) in the same energy range relevant for a light DM search. Hence for larger exposures the sensitivity of Direct Detection experiments is restricted to numbers of DM events larger than the Poisson uncertainty on the irreducible neutrino background, causing the limiting cross section to scale only as $\sigma_{\mathrm{lim}} \sim (\mathrm{exposure})^{-1/2}$.

For DM masses around 6~GeV the situation is particularly stark due to the close similarity between the recoil spectra from solar neutrinos and DM~\cite{Billard:2013qya}. Hence essentially no discrimination is possible between the recoil spectra of 6~GeV DM and the neutrino background and so any signal smaller than the systematic uncertainties on the $^8$B solar neutrino flux can not be observed or excluded. This leads to an effective `neutrino floor' where the sensitivity of Direct Detection experiments to $\sim$6~GeV DM improves only slowly with exposure beyond~$\sigma \sim 10^{-45}$~cm$^2$.

Fortunately there may be hope in the form of an additional discrimination parameter: the time-dependence of the DM and neutrino signals (see ref.~\cite{Grothaus:2014hja} for a related discussion based on directional detection). Both the solar neutrino\footnote{We focus on the competition between light DM and solar neutrinos in this work however there is also a neutrino background for heavier DM due to atmospheric and diffuse supernovae neutrinos~\cite{Billard:2013qya,Monroe:2007xp}. The rates of the latter are likely too low to give enough statistics in future experiments to make discrimination based on timing-information possible.} and DM signals are expected to vary by a few percent over a year, however the former is largest at the start of the year while the latter peaks some time around early June~\cite{Davis:2014cja}. In this work we seek to understand to what extent the different time-dependence of the two signals can be used to beat the `neutrino floor' and improve the sensitivity of Direct Detection experiments. 

In section~\ref{sec:shm} we discuss the case for DM within the Standard Halo Model (SHM) i.e. the commonly-taken assumption that the DM velocities are distributed in the galaxy according to a Maxwell-Boltzmann distribution. However in section~\ref{sec:not_shm} we extend our analysis to other forms of this distribution $f(v)$, as motivated by N-body numerical simulations of DM. We show that since $f(v)$ is not well-known this has implications for the ability of experiments to discriminate between the DM and neutrino signals based on their spectra. The uncertainty in the DM spectra means the neutrino-floor is extended to a wider range of DM masses, and we show using Bayesian statistics to what extent timing information can be used to surpass this limit.

\section{Improving the sensitivity of experiments to Dark Matter within the Standard Halo Model \label{sec:shm}}
\subsection{Expected signals from Dark Matter and neutrinos \label{sec:signals}}
\begin{figure}[t]
\centering
\includegraphics[width=0.97\textwidth]{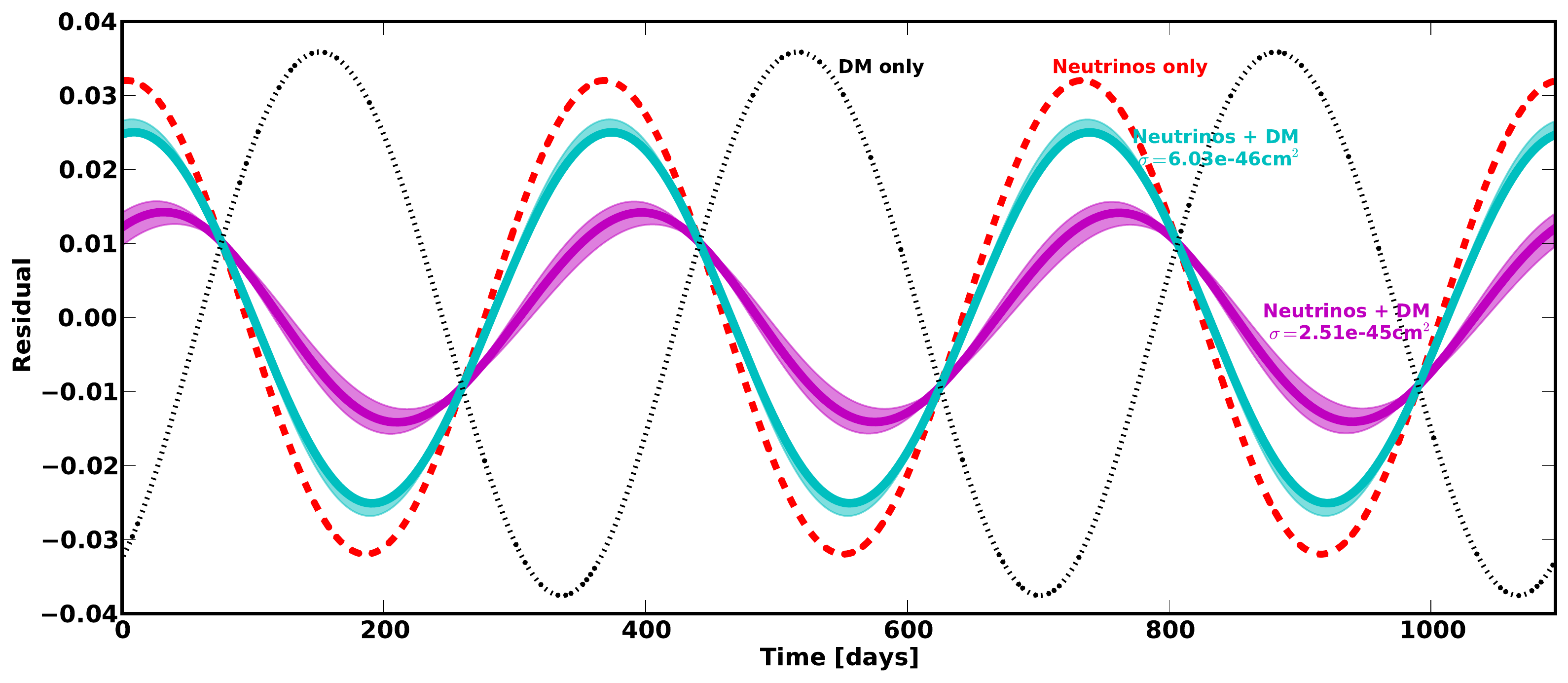}
\caption{Time variation from DM with a mass of 6~GeV when combined with the signal from solar neutrino induced nuclear recoils, summed over all recoil energies. \emph{Combining the DM and neutrino signals gives a new time-varying cosine with a different phase and amplitude.} A larger DM-nucleon cross section $\sigma$ causes the phase of the combined signal to shift to later times and reduces the size of the overall residual. The shaded regions indicate uncertainty due to systematics in the neutrino flux.}
\label{fig:combined_time}
\end{figure}

Both the DM and solar neutrino event rates are expected to vary over the course of a year\footnote{The DM signal should also possess higher-order modes with periods of two years and longer~\cite{Lee:2013xxa,Lee:2013wza,Bozorgnia:2014dqa} and also a monthly-modulation~\cite{Britto:2014wga}. We neglect these possibilities in our work for simplicity, though in principle they could be used to further improve the power of time-series data.}, however with different phases and residuals (defined as the fractional deviation of the modulated rate from the time-average). For the latter the modulation is due to the annual variation in the Earth-Sun distance, which changes by a few percent over the year due to the Earth's eccentric orbit.
The flux of solar neutrinos obeys an inverse-square relation: including the annual variation in the Earth-Sun distance it is given by~\cite{Bellini:2013lnn},
\begin{equation}
\Phi_{\nu} (t,E_{\nu}) = \frac{\mathcal{R}(E_{\nu})}{4 \pi r^2(t)} \approx \frac{\mathcal{R}(E_{\nu})}{4 \pi r_0^2} \left[ 1 + 2 \epsilon \mathrm{cos} \left(\frac{2 \pi (t - \phi_{\nu})}{T_{\nu}}\right)  \right] ,
\label{eqn:nu}
\end{equation}
where $\mathcal{R}$ is the neutrino production rate in the Sun, $t$ is the time from January 1st, $r(t)$ is the time-dependent distance between the Earth and Sun, $r_0$ is the average distance, $\epsilon = 0.01671$ is the orbital eccentricity, $T_{\nu}$ is the period and $\phi_{\nu}$ is the phase. The Earth is closest to the Sun around January 4th (implying $\phi_{\nu} = 3$~days).
Measurements from Borexino~\cite{Bellini:2013lnn} for $^7$Be neutrinos give a period of $T_{\nu} = 1.01 \pm 0.07$~years and a phase of $\phi_{\nu} = 11.0 \pm 4.0$~days and observations of $^8$B neutrinos by Super-Kamiokande~\cite{PhysRevD.78.032002} are consistent with equation (\ref{eqn:nu}). Hence the differential neutrino rate per unit detector mass is
\begin{equation}
\frac{\mathrm{d}^2 N_{\nu}}{\mathrm{d} t \mathrm{d}E} = \int \mathrm{d} E_{\nu} \, \sigma(E,E_{\nu}) \Phi_{\nu} (t,E_{\nu}) ,
\end{equation}
where $E_{\nu}$ is the neutrino energy (as opposed to the nuclear recoil energy $E$) which we integrate out and $ \sigma(E,E_{\nu})$ is the neutrino-nucleus cross section~\cite{Billard:2013qya,Grothaus:2014hja,Monroe:2007xp}. In this work we consider only the $^8$B neutrinos as a background to DM searches as they dominate for the energy ranges relevant for a light-DM search.

In order to calculate the rate of DM recoil events, and their annual modulation, we need to know the distribution of DM velocities in the rest-frame of the Earth. The DM rate is expected to vary by a few percent with an approximately sinusoidal modulation due to the changing direction between the Earth and the DM over the year~\cite{Freese:2012xd}.
The differential rate of Dark Matter interactions with nuclei takes the form of
\begin{equation}
 \frac{\mathrm{d}^2 N_{\mathrm{DM}}}{\mathrm{d} t \mathrm{d}E} = \frac{\rho_{\mathrm{DM}}}{m_{\mathrm{N}} m} \int_{v_{\mathrm{min}}(E)}^{v_{\mathrm{esc}}} \mathrm{d}^3 v \frac{\mathrm{d}\sigma}{\mathrm{d}E} v f(\mathbf{v} + \mathbf{v}_E(t)),
\label{eqn:mean_speed}
\end{equation}
where $\rho_{\mathrm{DM}} = 0.3$~GeVcm$^{-3}$ is the DM density, $v_{\mathrm{esc}}$ is the galactic escape velocity, $v_{\mathrm{min}}(E)$ is the minimum velocity needed to give a recoil of energy $E$, $m_{\mathrm{N}}$ is the mass of the target nucleus, $m$ is the DM mass and $\mathrm{d}\sigma / \mathrm{d}E$ is the differential DM-nucleus cross section. We write the latter as 
${\mathrm{d}\sigma}/{\mathrm{d}E} = \sigma  A^2 {m_N F(E)}/{2 \mu_p^2 v^2}$, where $\sigma$ is the `zero-momentum' DM-nucleon cross section on which we set limits, $A$ is the atomic mass of the target,  $\mu_p$ is the DM-proton reduced mass and the function $F(E)$ is the nuclear form-factor~\cite{Lewin199687}.

The integral is over the galactic DM velocity distribution $f(\mathbf{v})$ boosted into the Earth's rest-frame by $\mathbf{v}_E(t)$. The time-dependence enters via this term, expressed as  $\mathbf{v}_E(t) = \mathbf{v}_0 + \mathbf{v}_{\mathrm{pec}} + \mathbf{u}_E(t)$, where $\mathbf{v}_0 = (0,220,0)$~kms$^{-1}$ and the peculiar velocity $\mathbf{v}_{\mathrm{pec}} = (11.1 \pm 1.2,12.2 \pm 2.0,7.3 \pm 0.6)$~kms$^{-1}$~\cite{2010MNRAS.403.1829S}. For the relative velocity between the Earth and the Sun $\mathbf{u}_E(t)$ we use the expression from refs.~\cite{McCabe:2013kea,Lee:2013xxa} and we assume an escape velocity of $v_{\mathrm{esc}} = 544$~kms$^{-1}$. We assume implicitly that the DM is distributed in a non-rotating (or slowly rotating) spherical halo. This is backed up by results from numerical N-body simulations, which find that the halo should rotate slowly at $\sim 10$~km/s~\cite{0004-637X-581-2-799,Bett:2009rn}. However if the DM is distributed in a co-rotating `dark disc'~\cite{Fan:2013tia}  then the annual modulation may be different.
For now we assume a Maxwell-Boltzmann distribution for $f(\mathbf{v})$. Specifically we take the velocity distribution to be~\cite{Freese:2012xd}
\begin{equation}
f(\mathbf{v}) = N \, \mathrm{exp} \left( - \frac{3 |v|^2}{2 \sigma_v^2} \right),
\label{eqn:shm}
\end{equation}
for $v \leq v_{\mathrm{esc}}$ (and zero otherwise). This is the so-called Standard Halo Model (SHM) where $\sigma_v = \sqrt{3/2} \, |v_0|$ and $N$ is a normalisation constant. In section~\ref{sec:not_shm} we study the effect of using different forms for $f(v)$.

Figure~\ref{fig:combined_time} shows the time variation expected for neutrinos and DM alone compared to a combined signal from neutrinos and DM, with a mass of 6~GeV, for two different cross sections (and recoil energies above $\sim 0.1$~keV). The combined neutrino and DM modulated signals results in a new approximately sinusoidal modulation with a different phase and residual.
There is one subtlety to this plot in that the DM modulation actually changes sign at low energy for light DM, causing it to have a phase close to that from solar neutrinos, while figure~\ref{fig:combined_time} has been obtained by summing over all energies above threshold. This phase shift is taken into account in our analysis-proper, however for light DM it occurs at too low an energy to have any significant effect.

What is clear is that a larger DM-nucleon cross section suppresses the modulation residual with respect to neutrinos-only, and shifts the phase to later times. This is how timing information can be used in principle to improve the sensitivity of Direct Detection experiments e.g. with 10 ton years of exposure and for DM with a mass of 6~GeV and a cross section of $\sigma = 6 \cdot 10^{-46}$~cm$^2$ the number of DM events is within the systematic uncertainty on the neutrino flux and is too similar in spectrum to the neutrino signal to be distinguished effectively, unless timing information is used. However a large amount of statistics is required to measure the modulation residual and phase with enough accuracy to discriminate different DM-nucleon cross sections. We show in section~\ref{sec:stats_shm} precisely how much data is required.

\subsection{Exclusion sensitivity with and without timing information \label{sec:stats_shm}}

\begin{figure}[t]
\centering
\includegraphics[width=0.47\textwidth]{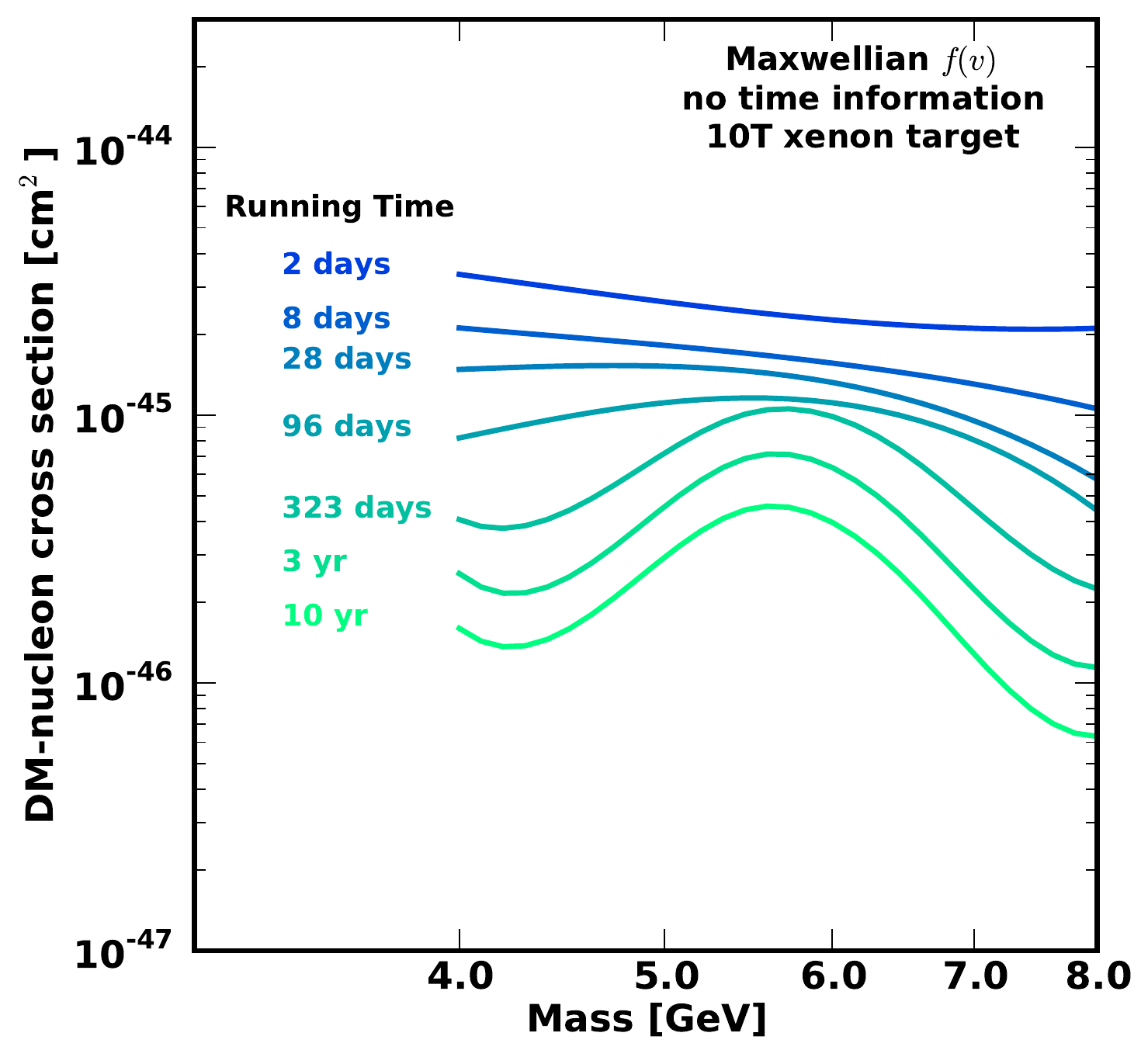} \hspace{10pt}
\includegraphics[width=0.47\textwidth]{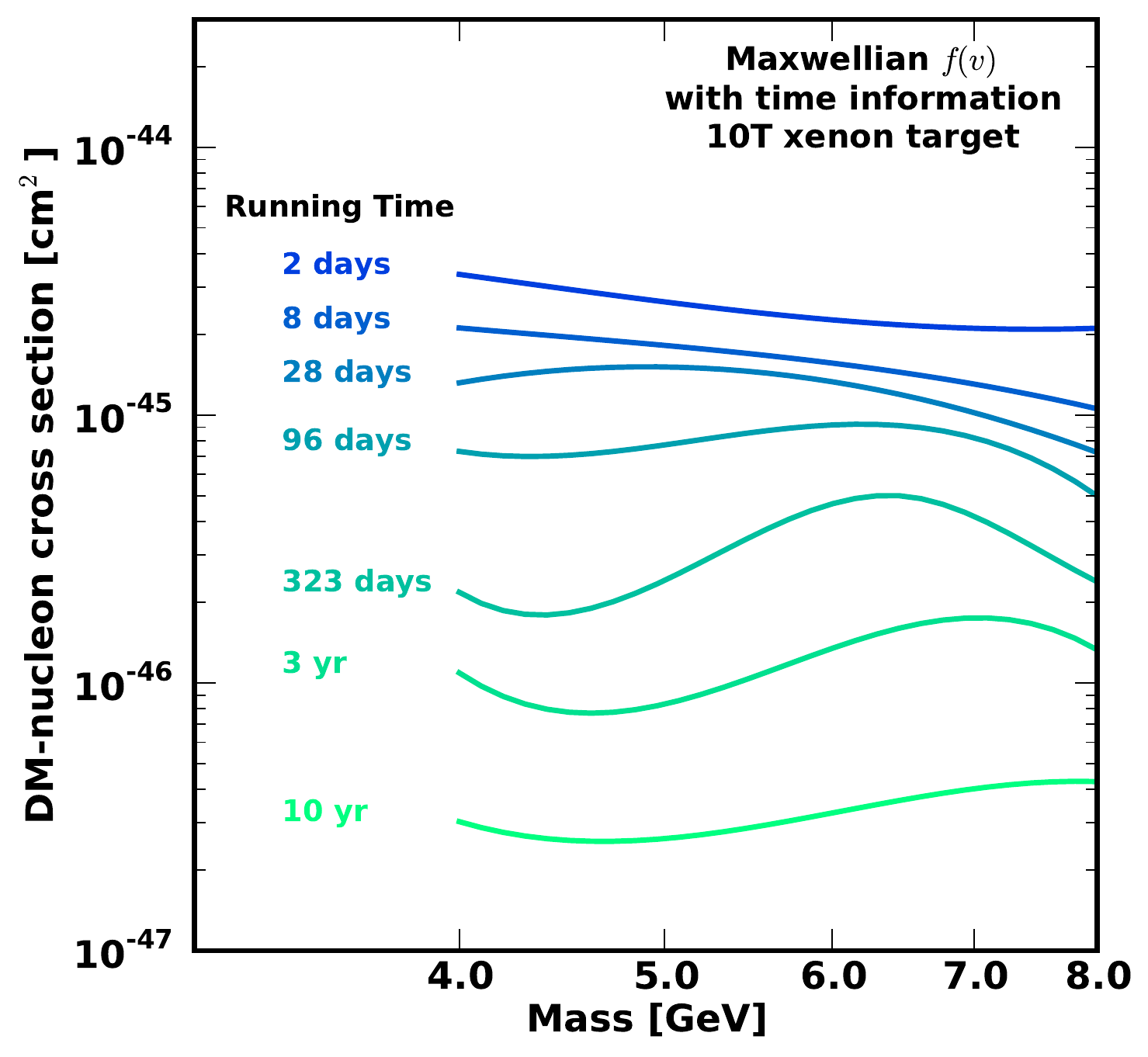}
\caption{Limits at $90\%$ confidence for DM within the Standard Halo Model using a hypothetical ten tonne low-threshold xenon detector at various running times. Without timing information (left) we can only use the different spectra $\mathrm{d}N / \mathrm{d} E$ to distinguish DM and neutrinos, resulting in a loss of sensitivity around 6~GeV where the spectra are almost identical. Adding timing information (right) results in improved limits as the DM and neutrinos are easier to tell apart.}
\label{fig:lims_shm_allm}
\end{figure}

In this section we seek to understand to what extent the neutrino background affects the ability of future Direct Detection experiments to exclude (or discover) light DM, and if using the different time-variation of the DM and solar neutrino signals (see figure~\ref{fig:combined_time}) can improve the sensitivity of these experiments. We perform two separate analyses: one including both time and spectral information and one with only the latter. For both analyses we treat the amplitude of the neutrino spectrum as a nuisance parameter, representing the uncertainty in the total $^8$B neutrino flux, and marginalise over it using a Gaussian prior. We assume a $15\%$ uncertainty on the total flux based on ref.~\cite{Billard:2013qya}, although smaller values have also been used~\cite{Grothaus:2014hja} (see e.g. refs.~\cite{Billard:2013qya,Grothaus:2014hja,Ruppin:2014bra} for the effect of smaller uncertainties on the neutrino floor). The details of the analysis are given in appendix~\ref{app1}. All of our limits are obtained using simulated data where we assume no DM signal i.e. only solar neutrino recoils.

In figure~\ref{fig:lims_shm_allm} we show the resulting $90\%$ confidence exclusion limits for a hypothetical ten tonne xenon experiment with a low-energy threshold of $0.1$~keV at several different running times\footnote{The level of the limit is only an approximate projection for future detectors. It is sensitive to how the limit is set (our Bayesian method differs from the one used in refs.~\cite{Billard:2013qya,Grothaus:2014hja,Ruppin:2014bra}) and the exact shape of the DM spectrum, which depends on particle and astro physics assumptions.}. We have assumed an idealised experiment with e.g. perfect neutron shielding such that all the backgrounds are from neutrino nuclear-recoils.
Without timing information the limit is significantly weaker around a DM mass of 6~GeV than at other masses, due to the similarity between the DM and neutrino nuclear-recoil spectra. Indeed the systematic uncertainties in the neutrino flux limit improvements in the exclusion sensitivity around 6~GeV masses. The limit improves again with $\sim 3$~years or more of running as there is enough statistics to exploit the small differences in the DM and neutrino spectra (in agreement with results from refs.~\cite{Ruppin:2014bra,Grothaus:2014hja}). However with timing information the limit is much stronger, especially around $\sim 6$~GeV, and {the limit is close to the value expected when the uncertainties are dominated by Poisson statistics, after 10 years of running}. Note though that timing information only improves the exclusion sensitivity with $\sim 1$~year or more of running, corresponding to approximately $10^4$ recoil~events.

{There are two more subtle points which are demonstrated in figure~\ref{fig:lims_shm_allm}. The first is that the DM mass at which the exclusion sensitivity is weakest, when using timing information, shifts to larger values with longer running time. This is because the DM mass which is most closely `mimicked' by the simulated neutrino data moves to larger values with longer running times.}
{The second is the non-trivial dependence of the exclusion sensitivity on running time. We demonstrate this more clearly for DM with a mass of 6~GeV in figure~\ref{lims_6}. In agreement with previous results~\cite{Billard:2013qya,Grothaus:2014hja,Ruppin:2014bra} the exclusion limit improves with $(\textrm{exposure})^{-1/2}$ when the Poisson errors on the neutrino background are the dominant source of uncertainty, and saturates at a roughly constant value once the systematic uncertainties on the neutrino flux start to dominate. Once there is enough data for the timing information to take effect the limit improves markedly, and indeed faster than $(\textrm{exposure})^{-1/2}$ around one year of running, as shown in figures~\ref{fig:lims_shm_allm} and~\ref{lims_6}. This rapid improvement is possibly due to a contribution from two factors: a purely statistical improvement due to an increased amount of data (scaling as ($\textrm{exposure})^{-1/2}$)  and an additional multiplicative contribution from the larger time-window over which the phase and amplitude can be measured\footnote{This may imply that up to one year of running changing the detector mass or running time does not have an identical effect on the derived upper limit, as is usually the case.}. Beyond this point the latter contribution has a negligible effect and the Poisson-scaling dominates, as shown by the dashed line in figure~\ref{lims_6}.

\begin{figure}[t]
\centering
\includegraphics[width=0.8\textwidth]{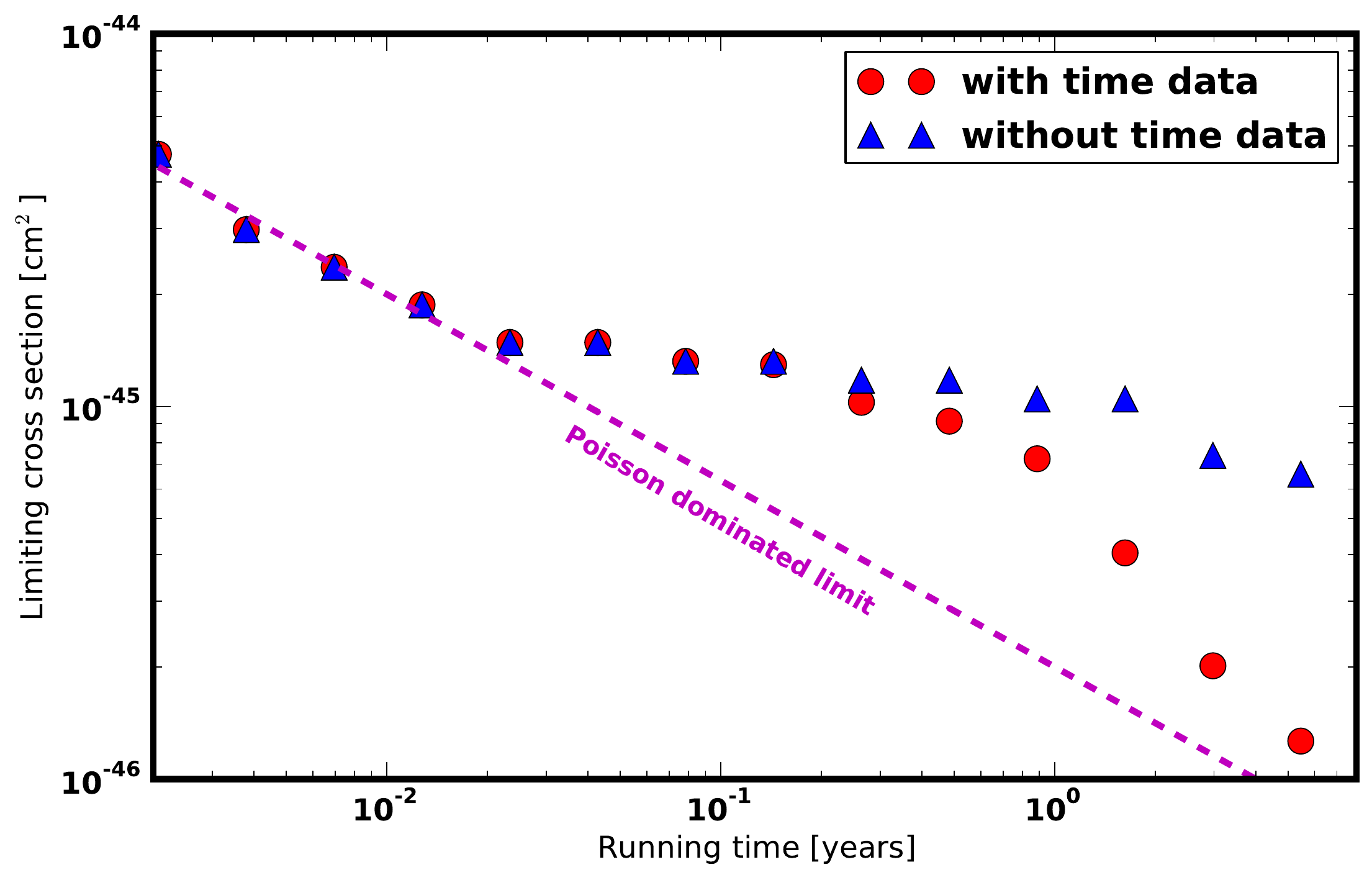}
\caption{Variation of the $90\%$ exclusion limit with running time for a hypothetical ten tonne xenon experiment, as in figure~\ref{fig:lims_shm_allm}, for DM with a mass of 6~GeV. The dashed line indicates the expected sensitivity when Poisson statistics represent the only source of uncertainty in the neutrino~background.}
\label{lims_6}
\end{figure}

A more simplified example of this behaviour is shown in figure~\ref{fig:lims_noE}. It shows the improvement gained from using timing information alone against using only the total number of events i.e. without any discrimination based on spectral information (hence why the limits are weaker than in figure~\ref{fig:lims_shm_allm}). Up until around $N_{\nu} \sim 10^4$ neutrino events the addition of timing information makes negligible difference to the exclusion limit. However above this value the exclusion power with timing information regains the $\sim (\mathrm{exposure})^{-1/2}$ dependence, while the limit without timing stays constant due to the systematics in the neutrino flux.  The fact that this transition occurs around $N_{\nu} \sim 10^4$ neutrino events can be understood by considering the statistical uncertainty on the event rate: assuming Poisson statistics the fractional uncertainty on the number of neutrino recoils is $N_{\nu}^{-1/2}$. Hence for $N_{\nu} \sim 10^4$ the fractional uncertainty is $\sim 1\%$ of the total rate. As can be seen from figure~\ref{fig:combined_time} we need precision at this level in order to actually distinguish different modulation scenarios based on their phase and residual, and so $N_{\nu} \gtrsim 10^4$ is required in order for timing-based discrimination to be viable. Note that the exclusion power in the presence of a large neutrino background depends also on the systematic uncertainties on the solar neutrino flux.

For larger exposures the exclusion power is limited only by the statistics required to measure the residual and phase. An example is shown in the right panel of figure~\ref{fig:lims_noE}: the filled region shows the phase and residual at $90\%$ confidence from fitting a simple cosine function with a period of one year to data from a hypothetical xenon experiment of mass ten tonnes with 10~years of running time.
We also show the phase and residual of the neutrino signal combined with DM with various cross sections. As expected from figure~\ref{fig:combined_time} a smaller DM-nucleon cross section results in a larger residual, closer to that from neutrinos alone, and an earlier phase. At $\sigma \approx 6 \cdot 10^{-46}$~cm$^2$ the residual and phase of the combined DM+neutrino signal are just outside the $90\%$ region and so can be excluded, even though the DM rate is too small to be observed above the neutrino background.
For smaller cross sections the statistics at this exposure is not good enough to exclude the signal based on timing information.

\begin{figure}[t]
\centering
\includegraphics[width=0.48\textwidth]{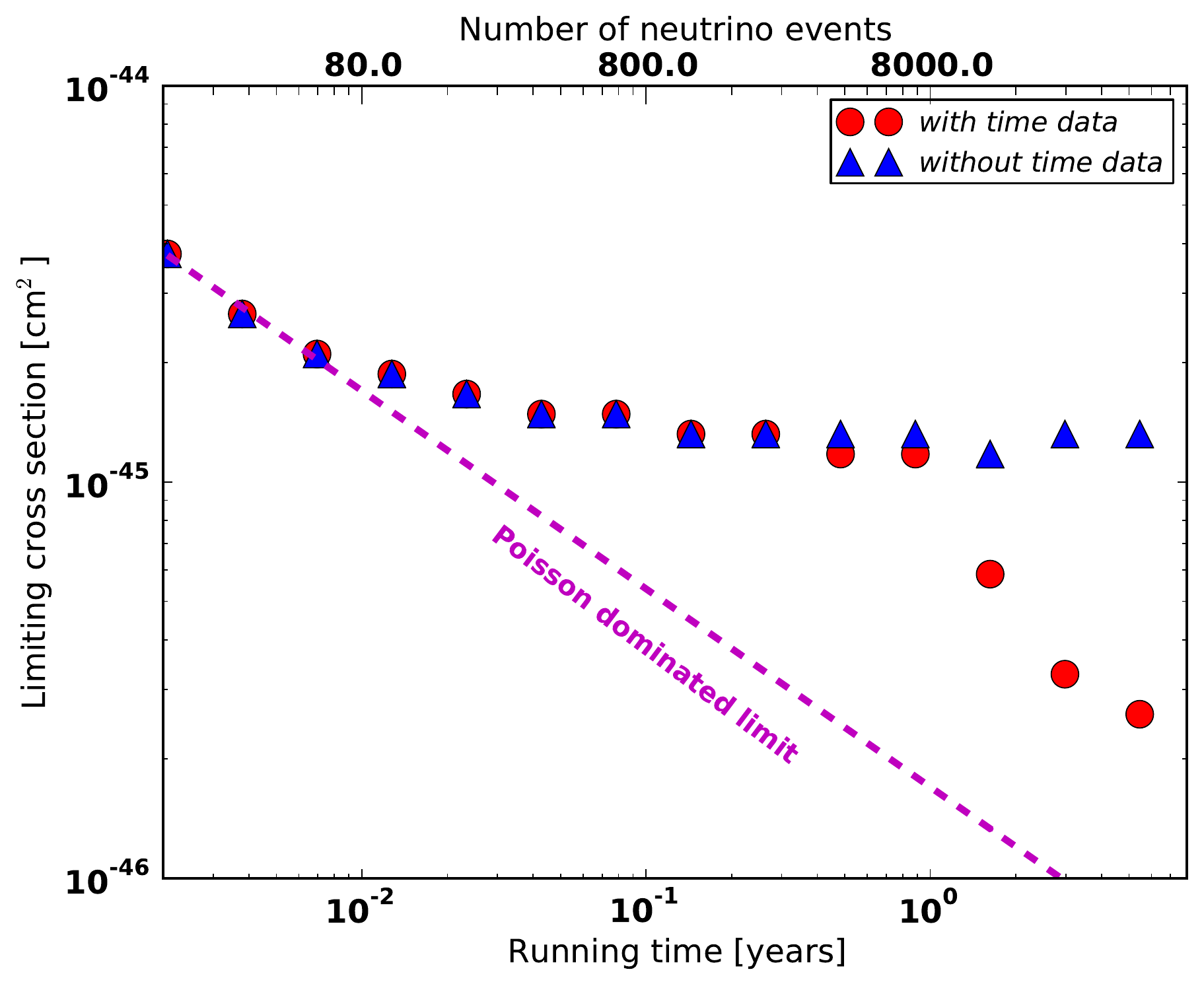} \hspace{5pt}
\includegraphics[width=0.48\textwidth]{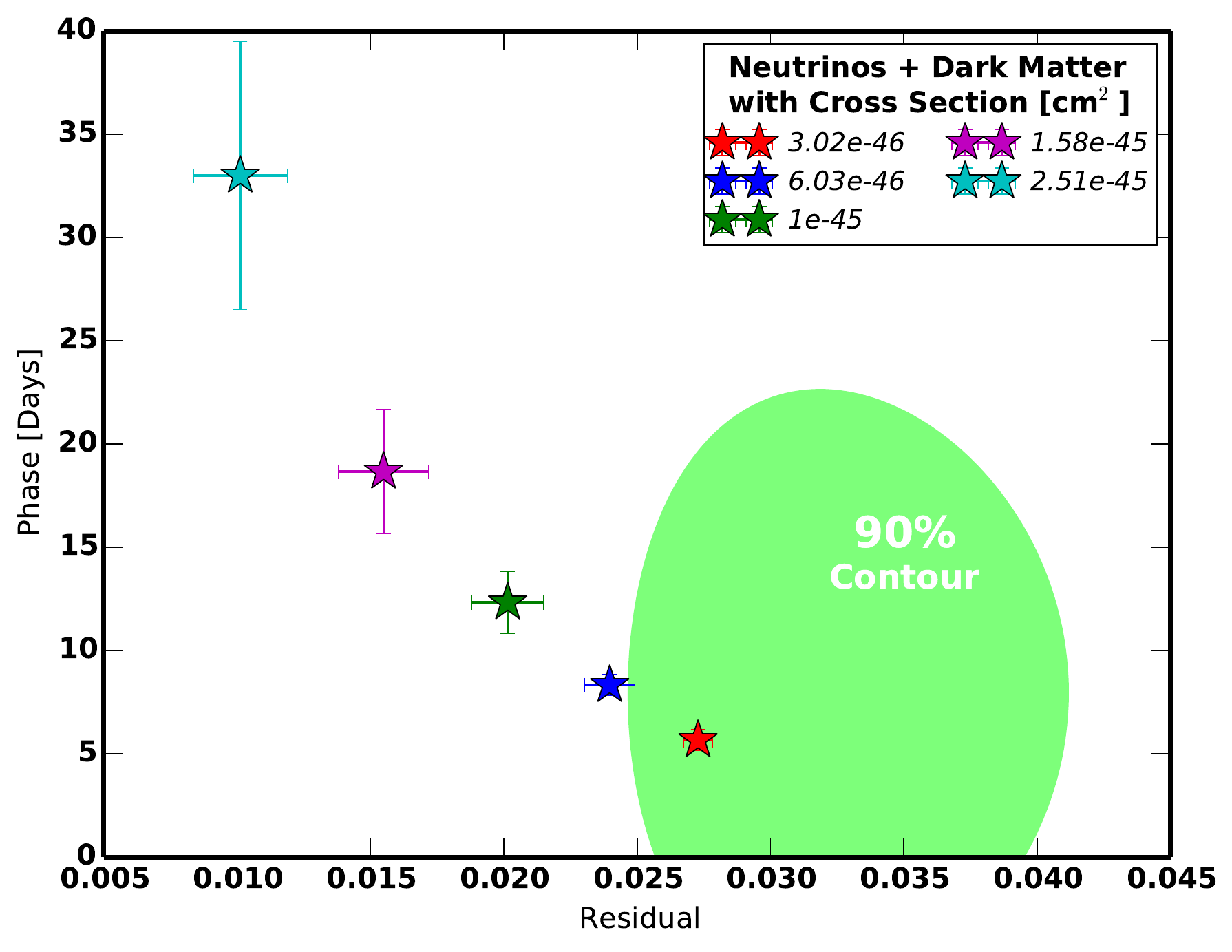}
\caption{\textbf{Left panel:} Example limits for a ten tonne xenon experiment with \emph{no spectral information} for 6~GeV mass DM. The dashed line indicates the expected sensitivity when Poisson statistics are the dominant source of uncertainty on the data.
\textbf{Right-panel:} With enough statistics ($\gtrsim 10^4$ events) the effective phase and residual of the combined DM and neutrino modulation can be used to exclude cross sections, even when the DM rate is too small to be observed directly. The shaded region gives the best-fit phase and residual at $90\%$ confidence when fitting to simulated data from 10~years of purely solar neutrino recoils.  The star points are the effective phase and residual of a combined DM and neutrino signal, with error bars arising from systematic uncertainties in the neutrino flux.}
\label{fig:lims_noE}
\end{figure}

Additionally we see that the effective residual of the combined modulated signal is a more powerful discriminator than the effective phase. Indeed using the phase of the combined signal alone would provide only a small improvement in sensitivity using timing information. Hence our conclusions are only valid if the solar neutrinos are the dominant nuclear recoil background for light DM, since an additional unmodulated background will also affect the residual of the combined signal.

\section{The effect on the neutrino background due to uncertainties in $f(v)$ \label{sec:not_shm}}
\subsection{Velocity distributions motivated by N-body simulations}

We showed in section~\ref{sec:shm} that a neutrino floor exists for DM masses close to 6~GeV due to the similarity between the DM and neutrino spectra, and that this can be surpassed to some extent using timing information. This was because, although their spectra are very similar, for most energies the annual modulation of DM and solar neutrinos is different in both phase and amplitude.
For other masses no such floor exists as the neutrino and DM recoil spectra are distinct enough to make discrimination possible despite the systematic uncertainties on the neutrino flux. Hence including timing information gave only marginal improvements in the expected exclusion sensitivity for $m \gtrsim 6$~GeV within the SHM.

This conclusion was reached under the assumption that the DM velocity distribution in the galaxy $f(v)$ is equal to a Maxwell-Boltzmann distribution i.e. the so-called Standard Halo Model (see equations (\ref{eqn:mean_speed}) and (\ref{eqn:shm})). However this is an over-simplification as $f(v)$ is poorly constrained~\cite{Kavanagh:2013wba,Feldstein:2014gza,McCabe:2010zh,Bozorgnia:2013pua,Fairbairn:2012zs,Freese:2012xd}. In this section we study the impact of using forms of $f(v)$ based on fits to results from numerical N-body simulations of Dark Matter in the galaxy~\cite{Mao:2013nda,Mao:2012hf}.
The DM recoil spectrum $ \frac{\mathrm{d}^2 N_{\mathrm{DM}}}{\mathrm{d} t \mathrm{d}E}$ depends crucially on $f(v)$ and so any uncertainties in this distribution will affect our ability to discriminate a DM signal from the neutrino background, which may extend the neutrino floor.

\begin{figure}[t]
\includegraphics[width=0.48\textwidth]{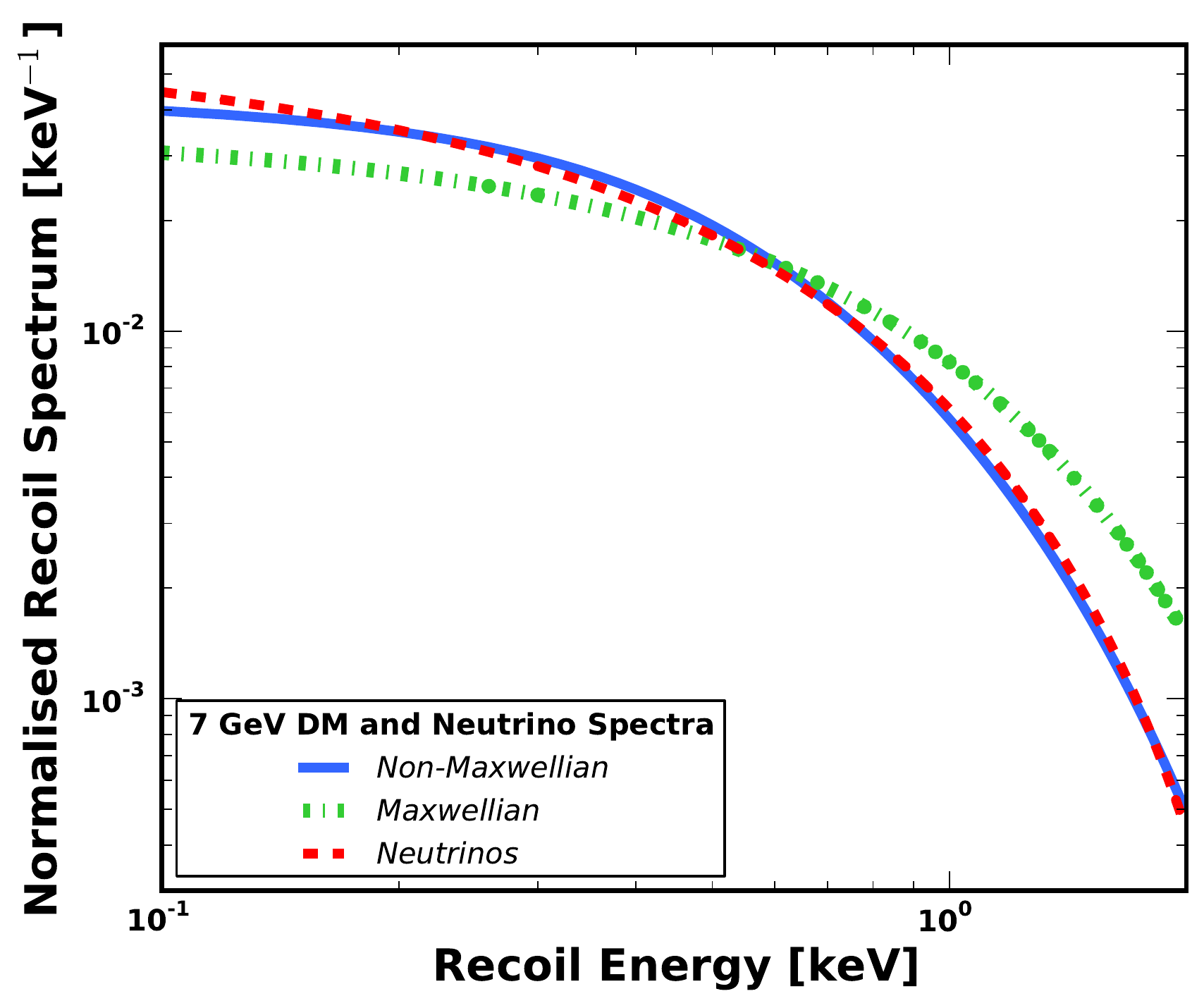} \hspace{5pt}
\includegraphics[width=0.48\textwidth]{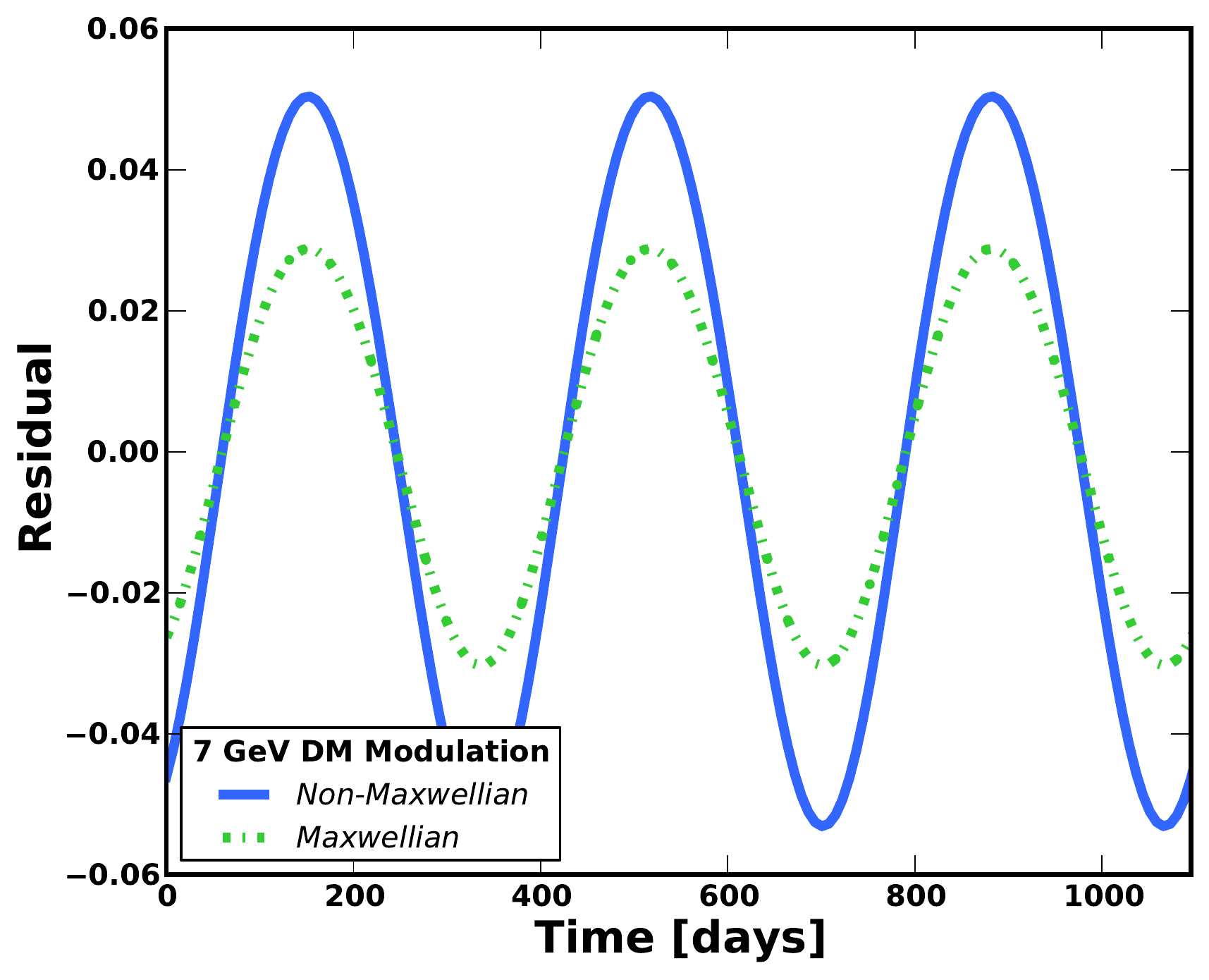}
\caption{\textbf{Left panel:} Spectra of nuclear recoils from Dark Matter with a mass of 7~GeV under two different assumptions for the velocity distribution in the galactic halo $f(v)$, compared with the expected spectrum from solar neutrinos. For the non-Maxwellian distribution we use the empirical formula from equation (\ref{eqn:mao_form}) with $p = 2$ and $v_{\mathrm{rms}} = 0.35 \, v_{\mathrm{esc}}$. \textbf{Right panel:} Annual modulation expected from the same Dark Matter models after summing over all recoil energies.}
\label{fig:best_DM_7}
\end{figure}

Hence in this section we perform our analysis using alternative forms for $f(v)$. We take for $f(v)$ the empirical formula
\begin{equation}
f(v) = A \, \mathrm{exp}(-v/v_0) (v^2_{\mathrm{esc}} - v^2)^p,
\label{eqn:mao_form}
\end{equation}
where $v$ is the velocity modulus and $f(v) = 0$ for $v > v_{\mathrm{esc}}$, and we define $A$ such that $4 \pi \int \mathrm{d} v \, v^2 f(v) = 1$.  This empirical formula is based on fits to numerical N-body simulations and so is likely to be a more physical assumption for the velocity distribution than the SHM\footnote{This is not the only empirical formula known to fit to results from N-body simulations \cite{Bozorgnia:2013pua,Fairbairn:2012zs,McCabe:2010zh,Green:2010gw,Mao:2012hf}. There are many alternatives, such as the Osipkov-Merritt model \cite{Bozorgnia:2013pua}, which allows for an anisotropic velocity distribution, or the Tsallis distribution \cite{Vergados:2007nc}.}.
In addition the uncertainty in this empirical fit is encapsulated in Priors (see appendix~\ref{app1} for a definition) for the parameters $p$ and $v_{\mathrm{rms}}$, which have been determined from fits to the \textsc{Rhapsody} and \textsc{Bolshoi} simulations to be within the ranges~\cite{Mao:2013nda,Mao:2012hf}
\begin{eqnarray}
p &\in& [0.0,3.0] \label{eq:p_prior} \\
v_{\mathrm{rms}} &\in& [0.35 \, v_{\mathrm{esc}},0.53 \, v_{\mathrm{esc}}] \label{eq:v_prior} ,
\end{eqnarray}
where the RMS velocity is defined as \cite{Mao:2013nda}
\begin{equation}
v_{\mathrm{rms}} = \left [4 \pi \int  \mathrm{d} v \, v^4 f(v) \right]^{1/2} .
\end{equation}
The extent of these Priors represents the large degree of variation between simulated DM halos seen in ref.~\cite{Mao:2012hf}. For our analysis we assume Priors which are constant within the ranges set by equations (\ref{eq:p_prior}) and (\ref{eq:v_prior}) and zero outside these ranges. We assume as before that $v_{\mathrm{esc}} = 544$~kms$^{-1}$, however in principle this value also depends on the velocity distribution and possesses a degree of uncertainty which would further broaden our Priors~\cite{Lavalle:2014rsa}. More details on our analysis can be found in appendix~\ref{app1}.

\begin{figure}[t]
\centering
\includegraphics[width=0.47\textwidth]{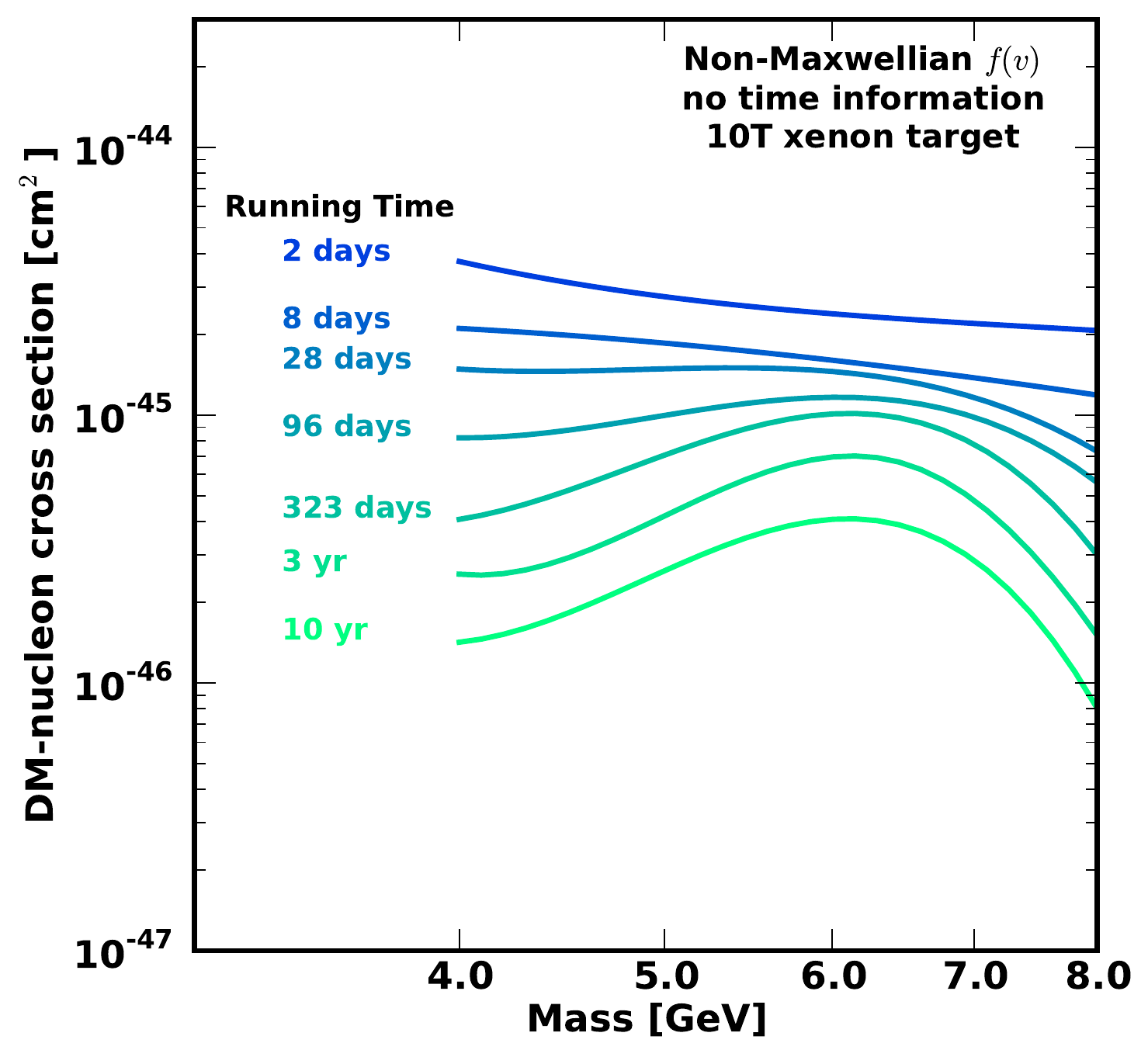} \hspace{10pt}
\includegraphics[width=0.47\textwidth]{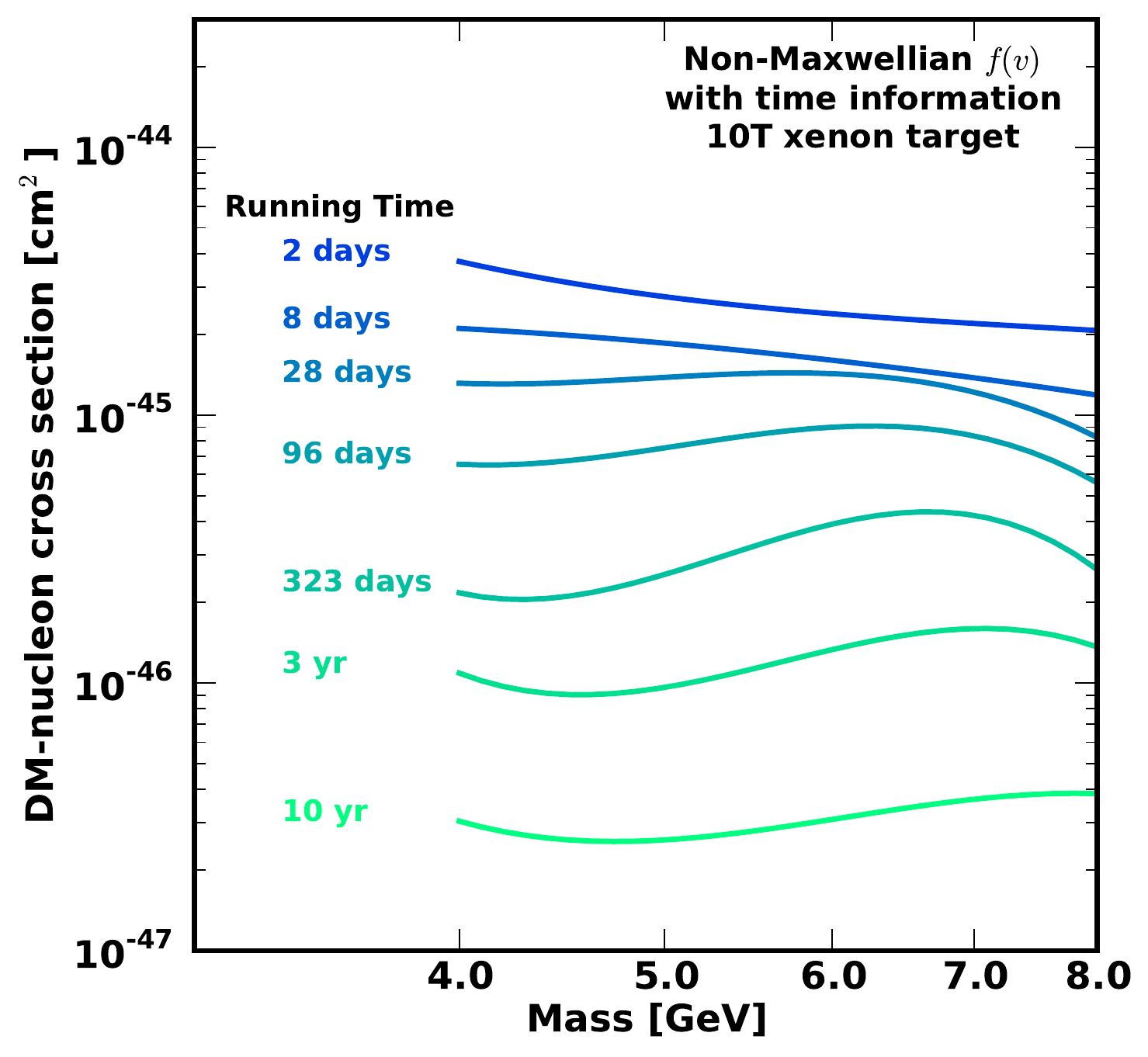}
\caption{Upper limits ($90\%$ confidence) at different running times for DM recoils under the assumption that their velocities are distributed according to equation (\ref{eqn:mao_form}) and marginalising over its free parameters. On the left panel we show exclusion limits using only the spectra of the neutrino and DM nuclear recoil signals, and on the right we use additionally the different temporal variation of the two signals as a further discrimination parameter.}
\label{fig:lims_mao}
\end{figure}

These uncertainties in $f(v)$ are passed on to the DM recoil spectrum $ \frac{\mathrm{d}^2 N_{\mathrm{DM}}}{\mathrm{d} t \mathrm{d}E}$ (equation (\ref{eqn:mean_speed})). For the current generation of Direct Detection experiments they have a relatively small effect on the exclusion power~\cite{McCabe:2010zh,Bozorgnia:2013pua}. However  this is no longer the case for future multi-ton experiments where the neutrino background becomes large. This can be understood from the left-panel of figure~\ref{fig:best_DM_7} where we show the DM recoil spectrum under a particular halo using $f(v)$ from equation~(\ref{eqn:mao_form}) and DM within the SHM, both for a mass of $7$~GeV, compared to the neutrino spectrum. For this particular choice of non-Maxwellian velocity distribution the DM recoil spectrum is almost indistinguishable from the neutrino spectrum, while the DM spectrum under the SHM is clearly distinct. Since we can not exclude the possibility that the DM has this particular form for $f(v)$, the inclusion of astrophysical uncertainties makes it considerably more difficult to distinguish between a signal from 7~GeV DM and neutrino nuclear-recoils.

As discussed in section~\ref{sec:stats_shm} the same issue arose for 6~GeV DM within the SHM resulting in much weaker limits being set, however 7~GeV DM was largely unaffected as its spectrum could be easily distinguished from the neutrino signal. Hence we expect that the increased uncertainties in the recoil spectrum using $f(v)$ from equation (\ref{eqn:mao_form}) will result in a broader neutrino-floor extended to 7~GeV masses (in addition to 6~GeV DM) and potentially beyond.

As can be seen from figure~\ref{fig:lims_mao} the uncertainties in the DM recoil spectrum due to marginalising over $f(v)$ using equation~(\ref{eqn:mao_form}) have extended the neutrino floor to heavier masses when using only spectral information, in line with our expectations.
Furthermore, as for the SHM, the neutrino floor is largely overcome by using timing-information, resulting in nearly an order of magnitude improvement for 6~GeV and 7~GeV after 10~years of running. This is not surprising as the DM annual modulation remains distinct from the solar neutrino signal even using $f(v)$ from equation~(\ref{eqn:mao_form}) (see the right-panel of figure~\ref{fig:best_DM_7}). As for the SHM about one year of running is needed to acquire enough statistics for timing information to improve the exclusion sensitivity, for a ten tonne low-threshold xenon detector. Hence the use of timing information is even more vital as it can surpass the neutrino floor even when the large uncertainties on $f(v)$ have extended it to heavier DM masses.

\subsection{Dark Matter velocity distributions with streams  \label{sec:not_shm_2}}

Up until now we have considered only smooth DM velocity distributions $f(v)$ i.e. those without significant features such as streams, which are effectively flows of DM particles moving with (approximately) the same velocity~\cite{Savage:2006qr,Lee:2013xxa,O'Hare:2014oxa}. Hence the temporal variation of the DM signal remained broadly unchanged despite large uncertainties in the recoil spectrum. However streams are known to change the temporal modulation of a DM signal~\cite{Savage:2006qr,Lee:2013xxa}, and so their presence could affect our ability to use timing information to discriminate between neutrino and DM recoils in future Direct Detection experiments.

As an example we consider the Sagittarius stream, which is the result of matter being accreted by our own galaxy from the Sagittarius dwarf galaxy~\cite{Savage:2006qr,Lee:2013xxa}. The accreted matter flow passes close to our own Solar System and should be accompanied by a stream of DM. The flow of DM from this stream has a direction different to that from the halo and we assume that its velocity is given by $\mathbf{v}_{\mathrm{Sgr}} = 330 \cdot (0,0.233,-0.97)$~kms$^{-1}$ (such that $\mathbf{v}_E(t) = \mathbf{v}_0 + \mathbf{v}_{\mathrm{pec}} + \mathbf{u}_E(t) - \mathbf{v}_{\mathrm{Sgr}}$) with a dispersion of $\sigma_{\mathrm{v}} = 25$~kms$^{-1}$~\cite{Savage:2006qr}, using the same formalism as in section~\ref{sec:signals}. We show the resulting time-variation of the DM recoil signal expected when this stream constitutes $10\%$ of the local DM density in the upper-panel of figure~\ref{fig:streams}, along with that from DM within the SHM and solar neutrinos.

\begin{figure}[t]
\centering
\includegraphics[width=0.98\textwidth]{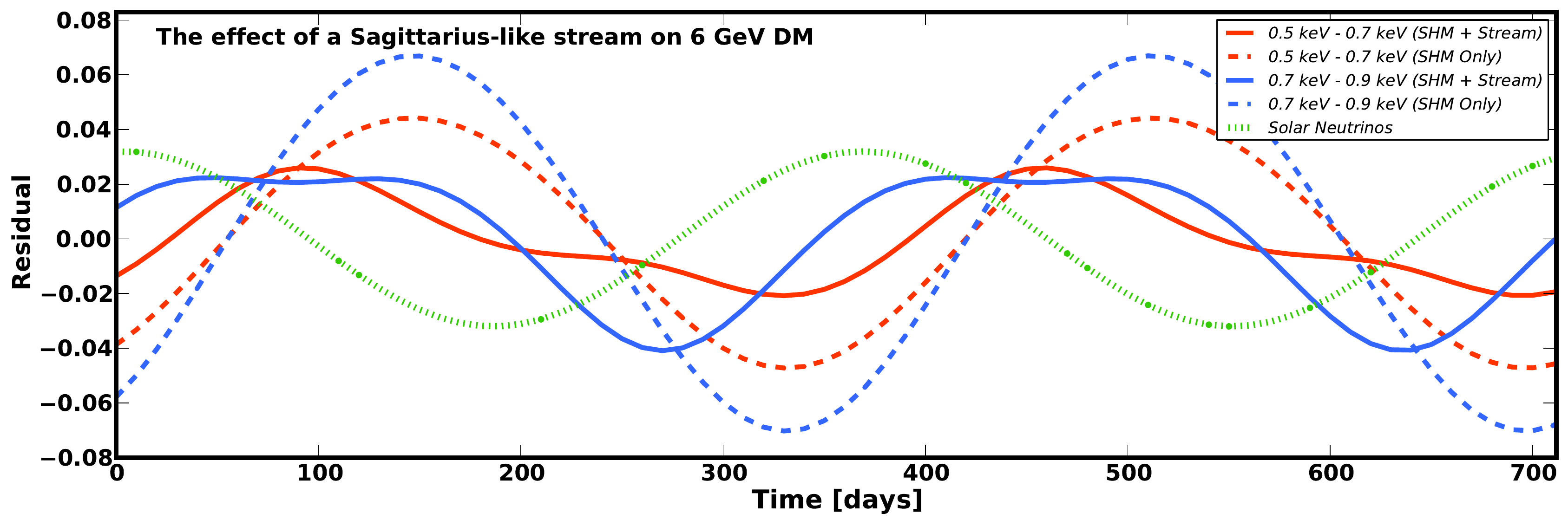} \\
\includegraphics[width=0.98\textwidth]{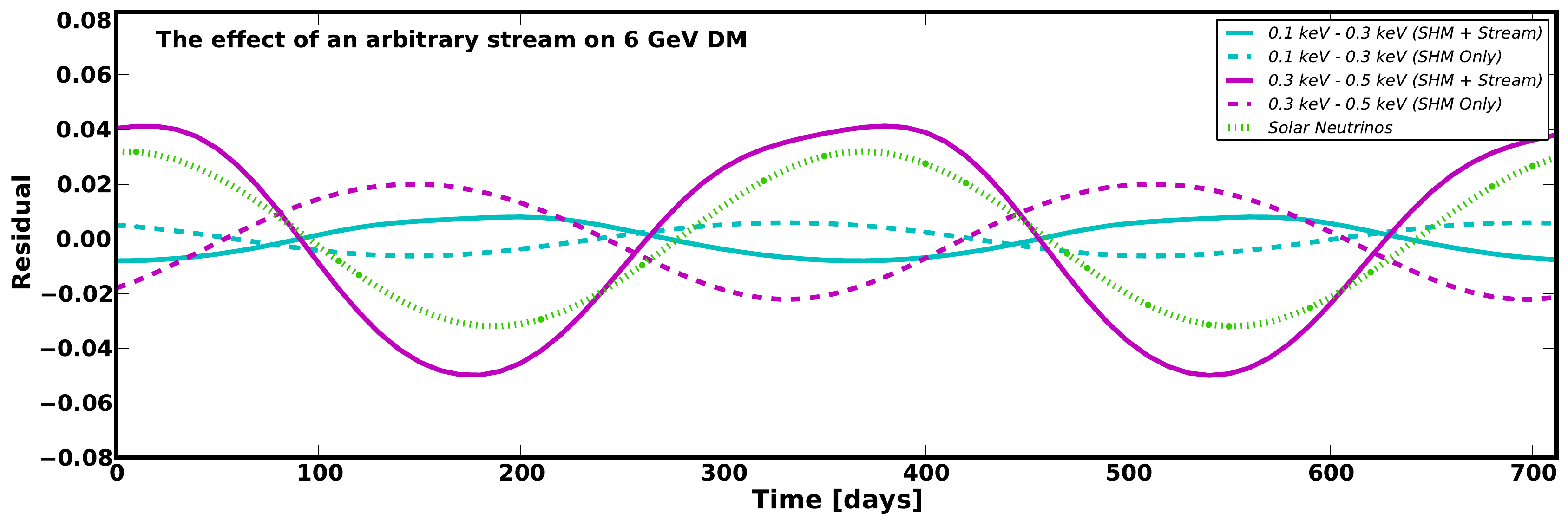}
\caption{Time variation expected from solar neutrinos and DM particles with a mass of 7~GeV for different recoil energy bins, under the assumption that their velocities are distributed according to a Maxwellian distribution alone or with an additional stream component. For this figure the stream constitutes $10\%$ of the DM density.}
\label{fig:streams}
\end{figure}

The effect of the stream is large only for recoil energies above $0.5$~keV and results in a modulation signal which departs significantly from a cosine and with a smaller maximum residual. The signal remains distinct from the solar neutrino modulation and so the time variation of the signal  is still an effective parameter for separating DM from neutrino recoils, even in the presence of a large Sagittarius-like stream.

A more interesting possibility is a stream whose time-variation is similar to that expected from solar neutrinos. We show in the lower-panel of figure~\ref{fig:streams} the time-variation expected for a stream with velocity $\mathbf{v}_{\mathrm{str}} = 250 \cdot (0,1,-1)$~kms$^{-1}$. For this stream the annually modulated DM signal at low-energy appears to a good approximation as a cosine with the opposite phase compared to the pure SHM case. Hence this time-signature may be more difficult to distinguish from the solar neutrino background, though higher energy recoils are largely unaffected.
As such the potential remains that particular velocity structures such as streams may be impossible to detect above the neutrino background, however combining timing and spectral information makes such a possibility very unlikely.

\section{Conclusion}
Future multi-tonne Direct Detection experiments face the difficult prospect of separating a potential Dark Matter nuclear-recoil signal from a large and irreducible neutrino recoil background~\cite{Grothaus:2014hja,Billard:2013qya,Ruppin:2014bra,Monroe:2007xp}. Specifically for light DM with masses up to $\sim 10$~GeV there is a significant background from $^8$B solar neutrinos. This limits the sensitivity of such experiments to numbers of DM events larger than the Poisson errors on the neutrino rate, leading to exclusion sensitivity scaling with the square root of the exposure rather than linearly, as has been true in the past.
For masses around $5$~GeV to $7$~GeV the improvement in exclusion/discovery sensitivity is even further limited due to the similarity between the DM and neutrino recoil spectra. This is the so-called `neutrino floor'. However this is relevant only when the neutrino and DM signals are separated based solely on their different recoil spectra. 

In this work we have examined how the different time-variation of the DM and neutrino signals can be used to surpass the neutrino floor, under various assumptions for the DM velocity distribution $f(v)$. Both the DM and solar neutrino rates vary over the year as approximate cosines, however the neutrino signal has a phase approximately 5~months earlier than for the DM annual modulation (except at very low recoil energies) and a different residual (the fractional deviation of the event rate from the time-average).
For DM-nucleon cross sections below the neutrino floor the time-varying DM and solar neutrino signals combine to give a new signal which retains the approximately sinusoidal variation.
However this has a later phase and reduced residual with respect to the solar neutrino signal, as for example shown in figure~\ref{fig:combined_time}. 

This is observable even when the DM spectrum $\mathrm{d} N / \mathrm{d} E$ is indistinguishable from the neutrino background.
\emph{Hence with enough statistics a future Direct Detection experiment can probe cross sections below the neutrino floor by measuring the phase and modulation residual of the time-dependent data.} Future experiments will need to observe $\gtrsim 10^4$ events to see any improvement, as with this many events the Poisson uncertainties are $\sim 1\%$ of the total rate i.e. around the same size as the fractional modulation amplitude for neutrinos and DM.
Indeed as shown in  figure~\ref{fig:lims_shm_allm}, for Dark Matter with a Maxwell-Boltzmann velocity distribution $f(v)$ i.e. the Standard Halo Model, after 10~years of running a ten tonne low-threshold (0.1~keV) xenon experiment can expect up to an order of magnitude improvement in exclusion power when using timing information as an additional discrimination parameter. Our conclusions should be qualitatively similar for other targets such as germanium.

Under more realistic assumptions for the velocity distribution $f(v)$, motivated by numerical simulations, and using only spectral information the neutrino floor is extended to heavier masses than using the Standard Halo Model (SHM), as can be seen by comparing the left panels of figures~\ref{fig:lims_shm_allm} and \ref{fig:lims_mao}. This is because the form of $f(v)$ is not well-known, leading to uncertainties in the DM recoil spectrum which make it harder to distinguish from the neutrino recoil spectrum, compared to the case when the SHM is assumed. This can be seen in the left-panel of figure~\ref{fig:best_DM_7}; {the spectrum for 7~GeV DM within the SHM is clearly distinct from the neutrinos, however for a different $f(v)$ they are almost identical}. Fortunately since the temporal variation of DM is still different from neutrinos (right-panel of figure~\ref{fig:best_DM_7}) the limit can again be improved significantly by using time information, as shown in figure~\ref{fig:lims_mao}, and the `neutrino floor' is essentially removed.

However this is still a fairly constraining assumption regarding the form of $f(v)$ i.e. that it corresponds to a smooth distribution from N-body simulations. More relaxed assumptions for the velocity distribution will result in larger uncertainties on the DM recoil spectrum and temporal modulation, which may broaden the neutrino floor further.
For example the presence of significant velocity streams may complicate matters as they can result in a time-varying DM signal closer in phase and residual to the solar neutrinos (see figure~\ref{fig:streams}). However this is generally only true for a small range of energies and so these streams will still be distinguishable from neutrino recoils when combining timing and spectral information. Even so this would be an interesting topic for further research.

\emph{Hence the `neutrino floor' is not an absolute limit on the sensitivity of Direct Detection experiments when timing information is used as an additional discrimination parameter between DM and neutrinos}. As such we find a similar result to previous work concerning for example the use of complimentary targets~\cite{Ruppin:2014bra} or directional information~\cite{Grothaus:2014hja} to overcome the neutrino background with additional discrimination parameters.
Even using timing information the sensitivity of future Direct Detection experiments is limited by the Poisson uncertainties on the neutrino background. Hence the strongest limit scales as $\sigma_{\mathrm{lim}} \sim (\mathrm{exposure})^{-1/2}$ (as shown in figure~\ref{lims_6}) and so the neutrino background can not be truly beaten without an event-by-event veto. Indeed to completely overcome this background multi-tonne experiments may have to move away from elastic DM-nucleus scattering, in order to distinguish DM and neutrino events more effectively.

\acknowledgments{The work of the author has been supported at IAP by  ERC project 267117 (DARK) hosted by Universit\'e Pierre et Marie Curie - Paris 6. The author acknowledges use of the high-performance computing facilities of IPPP, Durham University.}

\appendix
\section{Details of the Bayesian analysis \label{app1}}
Our limits have been derived using the Posterior $\mathcal{P}(d | \sigma,m)$, which is a function of the DM-nucleon cross section $\sigma$ and the DM mass $m$. For the case of the SHM in section~\ref{sec:shm} we use Bayes' theorem to relate this to the Likelihood function $\mathcal{L}(m,\sigma,A_{\nu};f(v)_{\mathrm{SHM}})$ as
\begin{equation}
\mathcal{P}(d | \sigma,m)_{\mathrm{SHM}} \, \mathcal{P}(d) = \int \mathrm{d} A_{\nu} \, \mathcal{L}(m,\sigma,A_{\nu};f(v)_{\mathrm{SHM}}) \mathcal{P}(m,\sigma) \mathcal{P}(A_{\nu}) ,
\label{eqn:mar_1}
\end{equation}
where for our purposes we can take the data-set Prior $\mathcal{P}(d) = 1$ and $A_{\nu}$ represents the systematic uncertainty in the solar neutrino flux. Since we do not want our final result to depend on our choice of $A_{\nu}$ we integrate it out, which means that we treat it as a nuisance parameter and marginalise over its values~\cite{Davis:2012hn}.
$\mathcal{P}(m,\sigma)$ and $\mathcal{P}(A_{\nu})$ are the Prior distributions for the DM parameters and the neutrino amplitude respectively. They represent the knowledge we have on $m$, $\sigma$ and $A_{\nu}$ before we perform our analysis. 

The form of the Likelihood function $\mathcal{L}(m,\sigma,A_{\nu};f(v)_{\mathrm{SHM}})$ depends on whether or not we are using both spectral and timing information or the former only. It takes the form of a Poisson distribution comparing theoretical expectation and data (which in this case is simulated) i.e.
\begin{eqnarray}
\mathcal{L}(m,\sigma,A_{\nu};f(v)_{\mathrm{SHM}}) = \begin{cases} \prod_{ij}  {\lambda_{ij}^{k_{ij}} e^{-\lambda_{ij}}}/{k_{ij}!} & \mathrm{with} \, \mathrm{timing} \, \mathrm{information} \\ 
\prod_{i}  {\lambda_{i}^{k_{i}} e^{-\lambda_{i}}}/{k_{i}!}  &  \mathrm{without} \, \mathrm{timing} \, \mathrm{information} \end{cases}
\end{eqnarray}
where $i$ denotes a particular bin in energy $E$, $j$ denotes a time $t$ bin and
\begin{eqnarray}
 \lambda_{ij} &=& M \int \limits_{E_i - \Delta E/2}^{E_i + \Delta E/2} \mathrm{d} E \int \limits_{t_j - \Delta t/2}^{t_j + \Delta t/2} \mathrm{d} t \left[ \frac{\mathrm{d}^2 N_{\mathrm{DM}}}{\mathrm{d} t \mathrm{d}E} (E,t;\sigma,m,f(v)_{\mathrm{SHM}}) + A_{\nu} \cdot \frac{\mathrm{d}^2 N_{\nu}}{\mathrm{d} t \mathrm{d}E} (E,t) \right]   \\ 
 \lambda_i &=& M \int \limits_{E_i - \Delta E/2}^{E_i + \Delta E/2} \mathrm{d} E \left[\frac{\mathrm{d} N_{\mathrm{DM}}}{\mathrm{d}E} (E;\sigma,m,f(v)_{\mathrm{SHM}}) + A_{\nu} \cdot \frac{\mathrm{d} N_{\nu}}{\mathrm{d}E} (E) \right] 
\end{eqnarray}
where $M$ is the detector mass and we have used that $\mathrm{d}N / \mathrm{d} E = \int \mathrm{d} t \, {\mathrm{d}^2 N} / {\mathrm{d} t \mathrm{d}E}$ over the running time of the experiment. Then $k_{ij}$ is the number of data points within the time values $t_j - \Delta t/2$ to $t_j + \Delta t/2$ and energies $E_i - \Delta E/2$ and $E_i + \Delta E/2$, and $k_i$ is the same value but integrated over all times $t$ for which the experiment has been running. In our analysis we use a time-bin size of $\Delta t = 10$~days and an energy-bin size of $\Delta E = 0.2$~keV, though our results are robust against reasonable changes in these values.

Since we know essentially nothing about the interactions of DM we take $\mathcal{P}(m,\sigma)$ to be constant. For the Prior on the neutrino amplitude $A_{\nu}$ we assume a Gaussian distribution such that
\begin{equation}
\mathcal{P}(A_{\nu}) = \exp \left[\frac{-(A_{\nu} - 1)^2}{2 \sigma_{\nu}^2} \right] ,
\end{equation}
where $\sigma_{\nu}$ is the fractional systematic uncertainty on the neutrino flux, which we have taken to be $\sigma_{\nu} = 0.15$ in this work.

For the non-Maxwellian velocity distribution $f(v;p,v_{\mathrm{rms}})$ used in section~\ref{sec:not_shm} we retain the same formalism as above, but with $f(v;p,v_{\mathrm{rms}})$ replacing $f(v)_{\mathrm{SHM}}$. In addition we have two new nuisance parameters $p$ and $v_{\mathrm{rms}}$ and so equation~(\ref{eqn:mar_1}) and our Posterior becomes
\begin{equation}
\mathcal{P}(d | \sigma,m) = \int \mathrm{d} A_{\nu} \, \mathrm{d} p \, \mathrm{d} v_{\mathrm{rms}} \, \mathcal{L}(m,\sigma,A_{\nu};f(v;p,v_{\mathrm{rms}})) \mathcal{P}(m,\sigma) \mathcal{P}(A_{\nu}) \mathcal{P}(p) \mathcal{P}(v_{\mathrm{rms}}) ,
\label{eqn:mar_2}
\end{equation}
where our Priors on $p$ and $v_{\mathrm{rms}}$ are defined as in section~\ref{sec:not_shm}.

We set exclusion limits for each value of $m$ by finding the cross section $\sigma_{\mathrm{lim}}$ at which $90\%$ of the Posterior is enclosed from above. Our limits are set using simulated `fake' data generated from the expected spectra and time-series of purely solar neutrino events, using the best-fit value of the neutrino flux (see e.g.~\cite{Grothaus:2014hja,Billard:2013qya,Ruppin:2014bra,Monroe:2007xp}). We generate $100$ independent data-sets $d_k$, where $k$ denotes a particular data-set, and combine these to set our projected limit. Hence we work with the Posterior marginalised over all such data-sets $\mathcal{P}(D | \sigma,m)$ given by (up to a normalisation factor)
\begin{equation}
\mathcal{P}(D | \sigma,m) = \sum_k \mathcal{P}(d_k | \sigma,m),
\end{equation}
where $\mathcal{P}(d_k | \sigma,m)$ is the Posterior for a particular simulated data-set $d_k$.
Hence our limit is defined using the expression
\begin{equation}
\frac{\int_0^{\sigma_{\mathrm{lim}}} \mathrm{d} \sigma \, \mathcal{P}(D | \sigma,m)}{\int_0^{\infty} \mathrm{d} \sigma \, \mathcal{P}(D | \sigma,m)} = 0.9 .
\end{equation}

\providecommand{\href}[2]{#2}\begingroup\raggedright\endgroup

\end{document}